\documentclass[%
 reprint,
 amsmath,amssymb,
 aps,
]{revtex4-1}

\usepackage{graphicx}% Include figure files
\usepackage{dcolumn}% Align table columns on decimal point
\usepackage{bm}% bold math
\begin{document}

%\preprint{APS/123-QED}

\title{Excitonic effects in two-dimensional TiSe$_2$ from hybrid density functional theory}% Force line breaks with \\

\author{Diego Pasquier}
 \email{diego.pasquier@epfl.ch}
\affiliation{Institute of Physics, Ecole Polytechnique F\'{e}d\'{e}rale de Lausanne (EPFL), CH-1015 Lausanne, Switzerland}

%\author{}
%\altaffiliation[Also at ]{Institute of Physics, EPFL}
\author{Oleg V. Yazyev}%
\email{oleg.yazyev@epfl.ch}
\affiliation{Institute of Physics, Ecole Polytechnique F\'{e}d\'{e}rale de Lausanne (EPFL), CH-1015 Lausanne, Switzerland}


\begin{abstract}
Transition metal dichalcogenides (TMDs), whether in bulk or in monolayer form, exhibit a rich variety of charge-density-wave (CDW) phases and stronger periodic lattice distortions.
While the actual role of nesting has been under debate, it is well understood that the microscopic interaction responsible for the CDWs is the electron-phonon coupling. 
The case of TiSe$_2$ is however unique in this family in that the normal state above the critical temperature $T_\mathrm{CDW}$ is characterized by a small quasiparticle bandgap as measured by ARPES, so that no nesting-derived enhancement of the susceptibility is present. 
It has therefore been argued that the mechanism responsible for this CDW should be different and that this material realizes the excitonic insulator phase proposed by Walter Kohn.
On the other hand, it has also been suggested that the whole phase diagram can be explained by a sufficiently strong electron-phonon coupling. 
In this work, in order to estimate how close this material is to the pure excitonic insulator instability, we quantify the strength of electron-hole interactions by computing the exciton band structure at the level of hybrid density functional theory, focusing on the monolayer.
We find that in a certain range of parameters the indirect gap at $q_{\mathrm{CDW}}$ is significantly reduced by excitonic effects.
We discuss the consequences of those results regarding the debate on the physical mechanism responsible for this CDW. 
Based on the dependence of the calculated exciton binding energies as a function of the mixing parameter of hybrid DFT, we conjecture that a necessary condition for a pure excitonic insulator is that its noninteracting electronic structure is metallic.

%\begin{description}
%\end{description}
\end{abstract}
%                             % Classification Scheme.
%%\keywords{Suggested keywords}%Use showkeys class option if keyword
\maketitle

%\tableofcontents

%\section{\label{sec:intro}Introduction}

%\section*{•} 
Transition metal dichalcogenides (TMDs) of chemical formula MX$_2$ (where M is a transition metal atom and X = S, Se, Te) are materials made up of layers weakly bond together by long-range van der Waals forces, each layer consisting of a triangular lattice of transition metal ions sandwiched between two layers of chalcogen atoms.
TMDs are realized in two polymorphs, depending on the coordination sphere of the transition metal atom that can have either trigonal prismatic or antiprismatic symmetry, leading to two families of materials called $2H$ (or $3R$ for bulk materials depending on the stacking sequence) and $1T$, respectively.
Periodic lattice distortions, often referred to as charge-density-waves (CDWs) when weak or moderate, are a recurrent phenomenon in both $2H$ and $1T$ metallic TMDs \cite{wilson_charge-density_1974, wilson_charge-density_1975, rossnagel2011origin, manzeli_2d_2017}. The CDW phases form rich phase diagrams and result in several interesting phenomena as side-effects such as anomalous metallic behaviour \cite{castro_neto_charge_2001}, topological phases \cite{qian_quantum_2014}, Anderson localization \cite{ovchinnikov2016disorder} or Mott insulating phases \cite{sipos_mott_2008, darancet_three-dimensional_2014, nakata_monolayer_2016} possibly associated with weak ferromagnetic correlations \cite{pasquier2018charge}.

TiSe$_2$ is known only in the $1T$ polymorph and belongs to the group IV TMDs, with a formal valence of $3d^0$ for Titanium.
The indirect quasiparticle gap in the normal state measured by ARPES is very small ($\sim 100$~meV in the monolayer \cite{chen2015charge, chen2016_dimensionality, sugawara2016_unconventional}) leading to the intensely-debated conjecture that the observed CDW with $T_{\mathrm{CDW}}\approx 200$~K is a manifestation of the excitonic insulator instability \cite{wilson1977concerning, wilson1978modelling, whangbo1992analogies, rossnagel2002_charge,cercellier2007_evidence, vanwezel2010_exciton, monney2012electron,  kaneko2018_exciton}, predicted several decades ago \cite{kohn1967_excitonic, jerome1967_excitonic} but still elusive in real materials (see however for instance Refs.~\onlinecite{varsano2017carbon, wakisaka2009_excitonic, kaneko2013_orthorombic}). 
This instability was originally proposed as a mechanism to interpret lattice distortions in small-gap semiconductors. 
Such an instability occurs when the electron-hole attraction generates an exciton binding energy larger than the bandgap. Notably, the semimetallic case with small band overlap is also prone to such an instability (see Ref.~\onlinecite{jerome1967_excitonic}). This leads to a particle-hole condensate associated with a distortion and breaks the translational symmetry of the crystal if the gap is indirect, due to the finite coupling between charge and lattice degrees of freedom \cite{monney2011_excitoncondensation}.
Another way of viewing the excitonic insulator phase is to notice that the exciton eigenenergies are the poles of the charge susceptibility renormalized by electron-electron interactions, so that a vanishing exciton energy would lead to an enhancement of its low-frequency part and therefore a soft phonon mode for a finite electron-phonon coupling. This is analogous to the original Peierls instability \cite{peierls1955quantum} where the \textit{bare} susceptibility is enhanced due to nesting properties \cite{chan1973spin}.

The difference between an excitonic insulator and a purely phonon-driven CDW is ultimately a quantitative one \cite{vanwezel2010_exciton, wakisaka2009_excitonic}, depending on the relative strength of electron-electron and electron-phonon interactions.
In any case the observable consequences are the same: a displacement of the atoms from their high-symmetry positions and a redistribution of the spectral weight, meaning that an electronic instability is challenging to disentangle. 
This problem is reminiscent of that of nematic ordering observed in iron-based superconductors, whose electronic origin remains difficult to demonstrate \cite{fernandes2014drives}.

\begin{figure*}[t]
%\centering
\includegraphics[width=17cm]{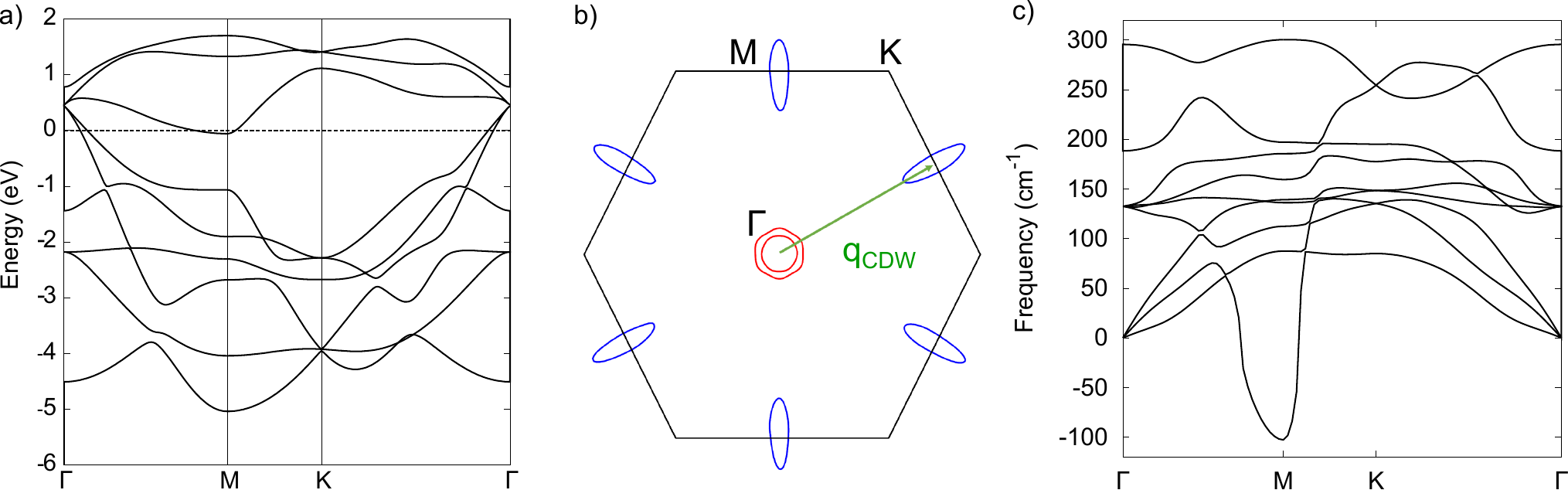}
\caption{\label{fig1} (a) Calculated electronic structure of monolayer $1T$-TiSe$_2$ at the PBE level. The dashed line corresponds to the Fermi level, set to zero. (b) Corresponding Fermi surface. The Selenium-derived hole bands are shown in red. The Titanium t$_{2g}$-like electron pockets are shown in blue. The CDW wave vector, shown in green, corresponds to the separation between the hole and electron pockets. (c) Calculated phonon dispersion at the PBE level along the high-symmetry directions. Imaginary frequencies are plotted as negative.}
\end{figure*}

Several experiments suggest important electron-electron effects in TiSe$_2$. 
In Ref.~\onlinecite{cercellier2007_evidence}, it was pointed out that the change in the spectral density measured by angle-resolved photoemission spectroscopy (ARPES), proportionally to the atomic displacements, is \textit{large} compared to other dichalcogenides.
More recently, using a new spectroscopic technique allowing to measure the momentum-resolved imaginary part of the charge susceptibility \cite{vig2017_measurement}, Kogar \textit{et al}. have found the evidence for a collective electronic mode softening just above $T_{\mathrm{CDW}}$ \cite{kogar2017_signature}.

On the computational side, Calandra and Mauri first showed that first-principles density functional calculations, within the local density approximation (LDA) and generalized gradient approximation (GGA), correctly  predict a soft phonon mode and reproduce a lattice distortion in fairly good agreement with experiments, casting serious doubts on the actual relevance of excitonic effects \cite{calandra_charge-density_2011, bianco2015_electronic}.
The drawback is that the normal state is wrongly predicted to be a metal with a Fermi surface displaying nesting between the hole and electron pockets at $\Gamma$ and $M$, leading to an artificial enhancement of the bare susceptibility at $q_{\mathrm{CDW}}$.
The insulating character can be recovered with the DFT+$U$ method, but the CDW is then lost as the gap opens \cite{bianco2015_electronic}.
A significant improvement was achieved by Hellgren \textit{et al}., who successfully reproduced the insulating character and distortion for bulk TiSe$_2$ within hybrid density functional theory, although the gap in the CDW phase was found to be overestimated \cite{hellgren_2017}.
It is unclear to what extent to proposed interpretation, i.e. an exchange-enhancement of the electron-phonon coupling, is distinct from an exciton-phonon driven instability, as the exchange interaction is known to be responsible for excitonic effects as well.

In this work, we provide an estimate of the strength of excitonic effects in this material from first-principles calculations, focusing on the monolayer.
The most general way to calculate exciton properties is to solve the Bethe-Salpeter equation with an exchange term screened by the full dielectric matrix, calculated in a preceding $GW$ calculation ($GW$+BSE) \cite{albrecht1998_abinitio}. 
Since there are no phonons involved in the problem, this should allow to determine whether a purely excitonic instability is indeed present.
This is unfortunately very demanding computationally, preventing us from obtaining converged results so far.
We have therefore adopted the method proposed by Kresse and collaborators \cite{paier2008_dielectric}, where the bare exchange in the BSE is screened more crudely in the fashion of hybrid functionals. 
Since this methodology is no longer fully \textit{ab initio}, we have varied the parameters in reasonable ranges and drawn ``phase diagrams''. 

\begin{figure*}[t]
%\centering
\includegraphics[width=17cm]{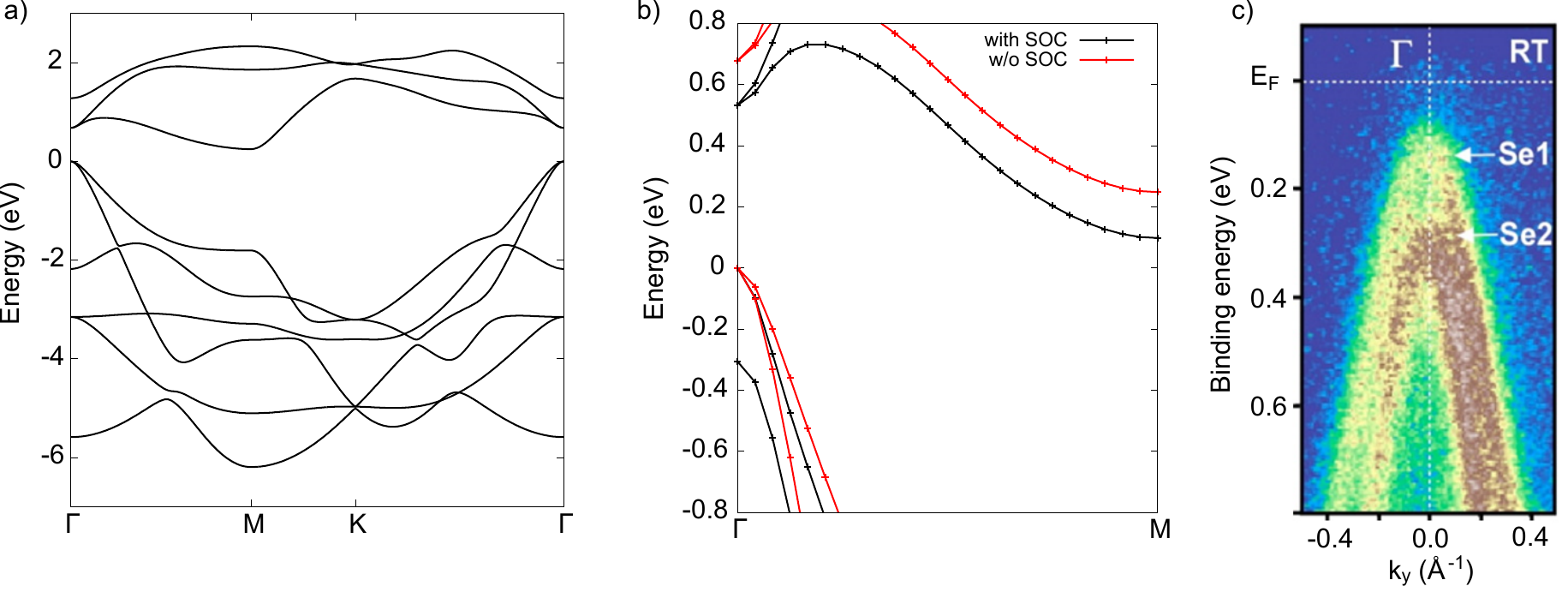}
\caption{\label{fig2} (a) Quasiparticle band structure along the high-symmetry directions using the PBE0 functional and $\alpha=0.185$. The valence band maximum is set to zero. (b) Calculated band structure, along the $\Gamma$-M direction, with and without the spin-orbit coupling. (c) ARPES data from Ref. \onlinecite{sugawara2016_unconventional} at room temperature along the $\Gamma$-M direction. The insulating character and spin-orbit induced splitting of the bands at the $\Gamma$ point are visible. 
%Reprinted with permission from Sugawara, Katsuaki, et al. "Unconventional charge-density-wave transition in monolayer 1 T-TiSe2." ACS nano 10.1 (2015): 1341-1345. 
Adapted with permission from Ref.~\cite{sugawara2016_unconventional}.
Copyright 2016 American Chemical Society. }
\end{figure*}

We begin by discussing the structural and electronic properties of the undistorted $1T$ phase at the GGA level \cite{methodology}, according to Perdew, Burke and Ernzerhof (PBE) \cite{perdew_generalized_1996}, using the \textsc{Quantum ESPRESSO} package \cite{giannozzi2009quantum} with PAW pseudopotentials \cite{blochl1994projector} from the pslibrary \cite{dal_corso_pseudopotentials_2014, pslibrary}.
This phase contains three atoms per unit cell and belongs to the $D^3_{3d}$ space group. 
The two independent structural parameters, namely the lattice constant $a$ and the Titanium-Selenium distance $d_{\mathrm{Ti-Se}}$, were calculated to be  $a=3.537$~\AA\  and $d_{\mathrm{Ti-Se}}=2.566$~\AA\, respectively. 
The calculated PBE lattice constant is in almost perfect agreement with the experimental value of $3.538$~\AA\ \cite{fang2017_xray}. 
This is not really surprising because, whereas the PBE functional is well-known to almost systematically overestimate lattice constants, it gives on average slightly underestimated lattice parameters for compounds with open $3d$ shells \cite{haas2009_calculation}.
The PBE electronic structure, shown in Fig.~\ref{fig1}a, is metallic, giving rise to a Fermi surface consisting of two hole pockets around the $\Gamma$ point and six electron pockets, derived from the Se $p$ and Ti $t_{2g}$ orbitals, respectively.
Following the work on bulk TiSe$_2$, several groups have recently reported the phonon softening and lattice distortion predicted by the PBE functional for the monolayer form \cite{singh2017_stable, wei2017_manipulating, guster2018first}.
In Fig.~\ref{fig1}c, we show the phonon dispersion, calculated using density functional perturbation theory \cite{baroni_phonons_2001}, displaying a soft mode at the $M$ point.
Like for the bulk case, the PBE prediction of the instability is associated with a \textit{metallic} electronic structure. 
Since imaginary frequencies in phonon calculations tend to be sensitive to the Fermi surface (e.g. charge doping suppresses the instability \cite{wei2017_manipulating, guster2018first}), it is unclear to what extent this prediction is robust or resulting partially from a cancellation of errors. 
We note that for TiS$_2$, PBE calculations predict a metallic electronic structure for both bulk and monolayer forms and phonon softening, with imaginary frequencies in the monolayer case \cite{dolui2016dimensionality}.
Experimentally, the bulk material exhibits a gap of $\sim 0.7$~eV \cite{suga2015momentum} and no sign of CDW or phonon softening were reported.

We now proceed to study the electronic and excitonic properties of monolayer TiSe$_2$ within hybrid density functional theory \cite{methodology}, using the VASP code \cite{kresse1993_abinitio, kresse1996efficiency, kresse1996efficient} with PAW pseudopotentials \cite{blochl1994projector, kresse1999_from}. 
For simplicity, the structure in the following calculations is obtained by relaxing the Ti-Se distance using the PBE+SOC functional with the experimental lattice parameter $a=3.538$~\AA. 
The most popular choices of hybrid functionals are the HSE and PBE0 \cite{adamo1999toward, heyd2003hybrid, krukau2006influence}, both mixing the PBE functional with $1/4$ of the Hartree-Fock exchange.
We have found that those two functionals, including the spin-orbit coupling (SOC), give quantitatively wrong electronic properties for this material. 
Indeed, the HSE06+SOC functional predicts a \textit{negative} gap of $\sim -0.1$~eV while the PBE0+SOC functional gives an overestimated gap of $\sim 0.4$~eV. 
These two functionals were designed to perform well for medium-gap semiconductors. 
The mixing parameter $\alpha$ can be interpreted as the inverse of the dielectric constant $1/\epsilon_{\infty}$ \cite{marques2011_density}, meaning that the chosen mixing parameter should be material-dependent. 
It has been shown that the PBE0 functional with mixing parameter determined self-consistently gives excellent accuracy in calculating bandgaps \cite{skone2014_selfconsistent}, justifying \textit{a posteriori} the use of $\alpha$ as a fitting parameter.

In Fig.~\ref{fig2}a-b, we show the electronic structure with the PBE0 functional with the mixing parameter $\alpha = 0.185$, with and without the spin-orbit coupling.
We see that the latter will play a crucial role in the following discussion as it reduces the gap by $\sim 0.2$~eV.
The lifting of degeneracy of the bands at the $\Gamma$-point is typical of $1T$ dichalcogenides and is due to the rather strong spin-orbit coupling coming from Selenium atoms.
We see in Fig.~\ref{fig2}c that this feature is also clearly observed in experiments.
The calculated band structure now correctly exhibits a small indirect quasiparticle bandgap between the top of the Se $p$ bands at $\Gamma$ and the bottom of the Ti $t_{2g}$ bands at the M point.

\begin{figure*}[t]
\includegraphics[width=17cm]{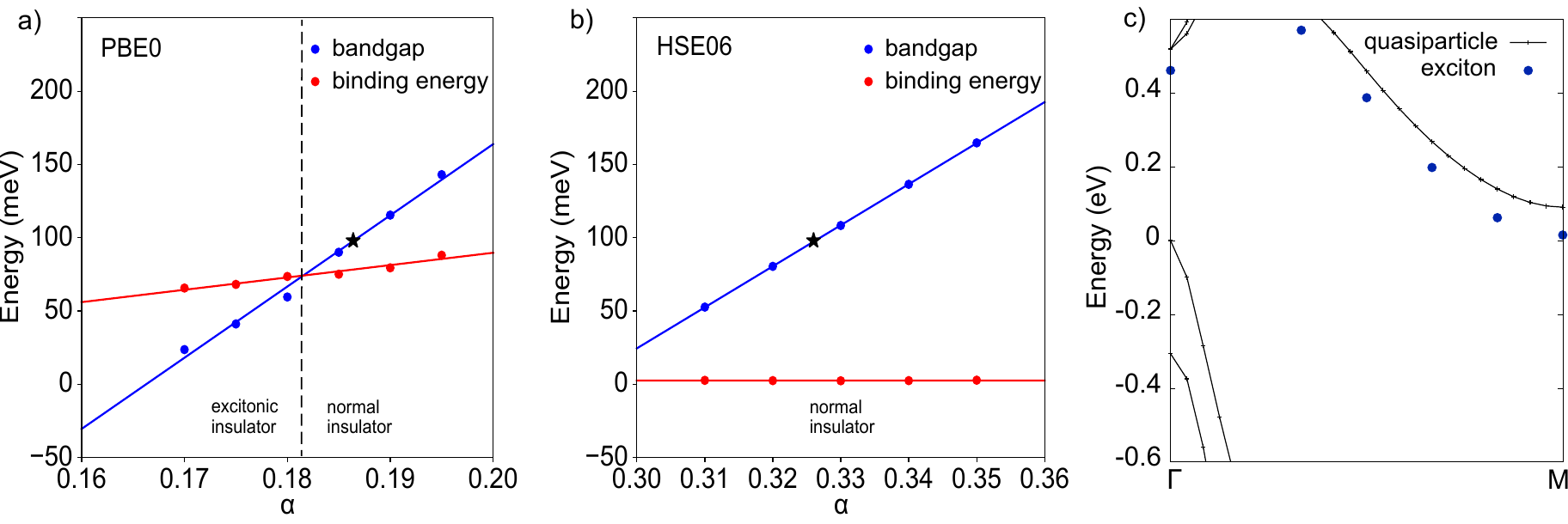}
\caption{\label{fig3} (a) Calculated indirect bandgap and exciton binding energy as a function of the mixing parameter $\alpha$ for the $\mu=0.0$~\AA$^{-1}$ functional (PBE0), including the spin-orbit coupling. The black star corresponds to the ARPES gap estimated at $98$~meV in Ref.~\onlinecite{chen2015charge}. The continuous lines drawn correspond to linear fits. The dashed line separates the normal and excitonic insulator phases. (b) Same as in (a), but with $\mu=0.2$~\AA\ $^{-1}$, corresponding to the HSE06 functional. (c) Band structure including the spin-orbit coupling for the mixing parameter $\alpha = 0.185$. The valence band maximum is set to $0.0$~eV. The blue dots correspond to the lowest-energy excitons for a few selected k-points.}
\end{figure*}

We then solved the Bethe-Salpeter equation at momentum $q_{\mathrm{CDW}}$ to obtain the finite-momentum exciton eigenenergies, to see whether the charge-neutral excitation gap is reduced compared to the quasiparticle one. 
The lowest-energy excitation corresponds to a dark exciton, i.e. with nearly zero oscillator strength, with an energy of $15$~meV, meaning an estimated binding energy of $75$~meV. 
Furthermore, we have repeated the following calculations for a series of mixing parameters, effectively varying the screening environment. 
In Fig.~\ref{fig3}a, we see that both the computed indirect quasiparticle gap and exciton binding energy at $q_{\mathrm{CDW}}$ scale linearly with the mixing parameter $\alpha$, although with different slopes.
We see that for smaller gaps, the two lines are crossing and the lowest-energy exciton has a \textit{negative} energy.
By fitting linearly the gap and binding energy, we estimate that the critical mixing parameter for an excitonic insulator is $\alpha_c \approx 0.181$, whereas the optimal parameter giving the ARPES gap of $98$~meV is $\alpha_{\mathrm{opt}} \approx 0.186$. 
We therefore conclude that within this approximation, the system is in proximity to a pure excitonic insulator instability. 

In Fig.~\ref{fig3}c, we plot the lowest-energy exciton for a few selected momenta along the $\Gamma$-M direction. 
One can see that the exciton band is roughly shifted by a constant value compared to the quasiparticle ones, showing that exciton binding energies are weakly momentum-dependent in this case.

We have also considered the effect of the screening parameter $\mu$ of the HSE functional \cite{heyd2003hybrid}. 
The latter is introduced in order to make the Coulomb potential short-ranged and is advantageous from a computational point of view as it facilitates the convergence with respect to the $k$-point mesh. 
We have therefore repeated the previous calculations by using $\mu = 0.2 \, \mathrm{\AA}^{-1}$, corresponding to the popular HSE06 functional \cite{krukau2006influence}. 
The effect on both the bandgap and excitonic properties is drastic. 
Indeed, the effect of the Fock self-energy on the gap is strongly reduced, so that a much larger mixing parameter $\alpha \approx 0.33$ is required to obtain the experimental gap. 
Moreover, we see in Fig.~\ref{fig3}b that excitonic effects are also strongly reduced. 
The exciton binding energies with this functional were calculated to be smaller than $3$~meV in the whole range of $\alpha$ considered, so that the system is now far from an excitonic insulator. 

To the best of our knowledge, systematic studies of how different hybrid functionals compare and perform for excitonic properties are missing, so it is difficult to tell \textit{a priori} which class of functionals gives the most reliable results, and the optimal choice is expected to be material-dependant.
We note that, in Ref.~\onlinecite{hellgren_2017}, it was found for the bulk material that the $\mu=0$ functional leads to the strongest enhancement of the electron-phonon couplings.
This correlates with our observation that the $\mu = 0$ limit gives the strongest excitonic effects.
This is rather natural, since the microscopic interaction leading to these effects is the same.
In a diagrammatic series, the dominant term in the BSE is the Fock diagram, which also appears in the expansion for the phonon propagator.
Hence, beside comparing our hybrid DFT result with the $GW$+BSE approach, it would be interesting to study the phonon dispersion with these two classes of functionals, but this is beyond the scope of the present work. 

Based on Fig.~\ref{fig3}a, we make the following observation. 
In order for the gap and binding energy to cross for some value of the mixing parameter $\alpha$, it seems necessary that the gap is negative at $\alpha =0$, since the attraction between electron and holes comes from the screened exchange and grows with $\alpha$ slower than the gap.
Therefore, it appears reasonable to conjecture that a necessary condition for a pure excitonic insulator is that its noninteracting band structure, i.e. without Fock self-energy corrections, is metallic.
Another observation we make is that while modeling  an excitonic phase using hybrid DFT is possible, a severe fine-tuning of the parameters is necessary.
We also stress that, as shown e.g. in Refs.~\onlinecite{vanwezel2010_exciton, kaneko2013_orthorombic}, a strong electron-hole interaction, even if not large enough to drive a purely electronic instability, reduces the minimal electron-phonon coupling necessary for a distortion to occur. 
The results obtained with the $\mu =0$ functional suggest such a scenario, in which both electron-phonon and excitonic interactions play important roles in the CDW phenomena.
It is clear that with a sizable electron-phonon coupling, the instability is much more robust and less fine-tuning is required.

In summary, we have investigated electronic properties and excitonic effects in monolayer TiSe$_2$ using hybrid density functional theory. 
Our calculations have revealed a region in the parameter space where the system is not far from a pure excitonic insulator, which would suggest a hybrid phonon-exciton mechanism. 
On the other hand, we have also found that the calculated ``phase diagram'' depends crucially on the choice of the screening length of HSE, with the choice $\mu = 0.2 \, \mathrm{\AA}^{-1}$ of HSE06 almost reducing to zero the exciton binding energy. 
The approximate and parameter-full character of hybrid DFT prevents us from drawing definite conclusions regarding a pure excitonic instability and we stress the desirability of comparison with higher-level theory. 
Nevertheless, the observed trends allow us to conjecture that a necessary condition for a pure excitonic insulator is that its noninteracting electronic structure is metallic.
\\

\section*{Acknowledgements}
We acknowledge funding by the European Commission under the Graphene Flagship (Grant agreement No.~696656). 
We thank Vamshi Katukuri and Igor Reshetnyak for discussions and Hyungjun Lee for technical assistance.
We also thank Katsuaki Sugawara for allowing us to reuse one figure.
First-principles calculations were performed at the facilities of Scientific IT and Application Support Center of EPFL and at the Swiss National Supercomputing Centre (CSCS) under project s832.
\bibliographystyle{apsrev4-1}
\bibliography{tise2}
\end{document}